\documentclass[twocolumn,showpacs,showkeys,superscriptaddress]{revtex4}

\usepackage{amsmath}
\usepackage{bbold}
\usepackage{amsfonts}
\usepackage{amssymb}
\usepackage{pbsi}
\usepackage[T1]{fontenc}
\usepackage{xcolor}
\usepackage{graphicx}
\usepackage{subfig}
\usepackage{float}
\usepackage{color}
\usepackage{hyperref}
\hypersetup{colorlinks = true,
	linkcolor = red,
	anchorcolor = blue,
	citecolor = blue,
	filecolor = blue,
	urlcolor = blue}

\begin{document}

\title{Electromagnetic plasma waves in dark energy cosmology}

\author{Felipe A. Asenjo}
\email{felipe.asenjo@uai.cl}
\affiliation{Facultad de Ingenier\'ia y Ciencias,
Universidad Adolfo Ib\'a\~nez, Santiago 7491169, Chile.}

\date{\today}

\begin{abstract}
 We explore  electromagnetic wave modes that can exist in a cosmological plasma dominated by dark energy due to a cosmological constant. It is found that, in the cold and hot plasma cases, electromagnetic plasma wave modes can be found exactly. The
effect of this cosmology appears as a time-dependent potential in the
wave equation for the electromagnetic modes, that effectively modify the frequency response of the plasma. This potential depends on the metric of the spacetime and on the thermodynamical properties of the plasma. For both cases, cold and hot, the solutions are found in terms of Airy and Bessel functions, respectively.  When those solutions are required to  have vanishing initial conditions, a discretization on the wavelengths of the electromagnetic plasma waves is imposed. Thus, only some specific wave modes can exist in this dark energy cosmology. Relaxing those conditions, we obtain other solutions that approximate to plane waves only in the very hot plasma limit.
\end{abstract}



\maketitle

\section{Introduction}

The expansion of the Universe affects the form in which electromagnetic waves propagate in a plasma. These cosmological electromagnetic plasma waves have different properties depending on which cosmological setting is developing \cite{book}.
The main effect induced by cosmology is the modification of the rate of time (to a conformal cosmological time) in which the wave phenomena evolve. This conformal time depends on the type of cosmology we are studying. For example, electromagnetic plasma waves in a radiation-dominated early-Universe acquire a Hankel form on its time oscillatory dependence, that only can be interpreted as a plane wave when time goes to infinity \cite{holcomb}.
This kind of richness in plasma waves (not always behaving as plane waves) is a consequence inherited from cosmology. Thus, many research have been carried out studying different plasma wave properties in different cosmologies (see for example, Refs.~\cite{Dodin,dettmann,tajfmie,Schlickeiser,lazar,Banerjee,Ramezani,Servin,Zimdahl,Sil,kosta,
gallis,Papadopoulos}, among others). 

It is usual to study plasma waves in a perturbative scheme. However, when the cosmological spacetime is a background to the plasma dynamics, the perturbation scheme is not necessary to study electromagnetic plasma waves.
These waves can be studied in an exact fashion, directly from Maxwell and plasma fluid equations, all of them written by considering a cosmological spacetime as a background metric. In this work, we perform such a study for a cosmology dominated by dark energy due to a cosmological constant. For this, we use the Magnetofluid unified tensor formalism \cite{mahajan,fazma,fazma2}, previously extended to cosmology in Ref.~\cite{faz}. Because of this dark energy cosmology, the conformal time determines specific and different solutions for cold and hot plasmas. They behave differently as time evolves because of their temporal dependence on the decrease of plasma temperature. Besides, this cosmological model
produces rules of discretizations on the possible wavelengths of the electromagnetic plasma waves when vanishing initial conditions are imposed. Thus, only certain waves can exist exactly in this dark energy cosmology. This is a very different behavior compared to the continuum spectra of usual electromagnetic waves in vacuum, even in cosmology. Thus, the plasma (to be specific the response plasma frequency) produces a major change in the form that an electromagnetic wave oscillates in the Universe.

Along this work, the dark energy cosmological model due to a cosmological constant $\Lambda$
is described by the isotropic flat Friedman-Lema\^itre-Robertson-Walker (FLRW) metric $g_{\mu\nu}={\mbox{diag}}(-1,a^2,a^2,a^2)$, written in cartesian
coordinates (with $\mu,\nu=0,1,2,3$), where the scale factor of the Universe is given by
\begin{equation}
    a(t)=\exp(H_0(t-t_0))\, ,
    \label{scalefactor}
\end{equation}
in terms of the cosmological time $t$, 
with $H_0=\sqrt{\Lambda/3}$ \cite{ryden}, and the current  time of the Universe $t_0$ (where $a=1$). This model considers only the energy density of the cosmological constant as  responsible for the accelerated expansion of the Universe. 

Besides, in the below calculations, we consider a non-gravitating two-fluid electron-positron plasma evolving in this cosmological background.  This implies that the only source of spacetime curvature is the cosmological constant (as in the case of cosmic inflation), as the plasma energy density is neglected in Einstein equations.
 In this case, each plasma fluid is described by the momentum equation
\begin{equation}
\nabla_\nu\left(h U^\mu U^\nu+p g^{\mu\nu}\right)=q n F^{\mu\nu}U_\nu\, ,
\label{eq1plasmasys}
\end{equation}
where  $U^\mu$ is the
normalized  four-velocity of the corresponding plasma fluid, and $\nabla^\mu$ the covariant derivative calculated from the metric $g_{\mu\nu}$. Besides,  $n$, $h$, and $p$ are the number density, the plasma density enthalpy, and the plasma pressure of the corresponding fluid. Lastly, $q$ is the electric  charge of the plasma fluid, and 
$F^{\mu\nu}$ is the electromagnetic tensor.

We can re-write Eq.~\eqref{eq1plasmasys} 
using the Magnetofluid unified tensor $M^{\mu\nu}=F^{\mu\nu}+(m/q) S^{\mu\nu}$ \cite{mahajan,fazma,fazma2}, where $m$ is  mass  of the electron,  and $S^{\mu\nu}=\nabla^\mu (f U^\nu)-\nabla^\nu (f U^\mu)$ is the antisymmetric fluid tensor. Also, $f=h/(mn)$ is a function of temperature
$T$,  that in the special case of a relativistic Maxwell distribution
becomes $f(T) = K_3(m/T )/K_2(m/T )$, where $K_j$ is the modified
Bessel functions of order $j$ (the Boltzmann constant is chosen
as unity) \cite{mahajan,fazma,fazma2,faz}. Thereby,  each plasma fluid dynamic described by the momentum equation \eqref{eq1plasmasys},  in terms of the unified tensor is written as
\begin{equation}
    U_\nu M^{\mu\nu}=0\, ,
    \label{magnetofluid1}
\end{equation}
for a homentropic plasma case, where $\partial_\mu p=m n \partial_\mu f$. Here, we refer the reader to Ref.~\cite{faz}, where specific details of this equation and its derivation appear. 

This electron-positron plasma fluid
is coupled to electromagnetic fields by Maxwell equations
\begin{equation}
    \nabla_\nu F^{\mu\nu}= e n \left(U_p^\mu -U_e^\mu\right)  \, ,
\label{ecMAxwellll}
\end{equation}
where $U_e$ ($U_p$) is the electron (positron) plasma four-velocity. Here,
$e$ is the magnitude of the electron charge, and we have assumed that both fluids have the same number density. From here, we obtain that each fluid fulfills the continuity equation
\begin{equation}
    \nabla_\mu\left( n \, U_{e,p}^\mu\right)=0\, .
\end{equation}

In the next section we show how the dark energy cosmology produce  different exact solution for electromagnetic plasma waves in the cold and hot case scenarios. Also, they produce discretization on the possible wavelengths of the electromagnetic waves. Finally, in Sec. III, we discuss these results.

\section{Electromagnetic plasma waves}

In an isotropic cosmology, the electromagnetic plasma wave equation can be found exactly 
for a two-fluid electron-positron plasma. 
In the homentropic case, the momentum equation \eqref{magnetofluid1} can be solved by  $M^{\mu\nu}=0$, implying that for each fluid we have
\begin{equation}
F^{\mu\nu}=-\frac{m}{q} S^{\mu\nu}\, .
\end{equation} 
Using the FRW cosmological metric, the above solution  for the electron plasma fluid can be put in the form \cite{faz}
\begin{equation}
   \nabla\times\left(f U_e^i \right)=-\frac{e a}{m} B^i\, .
\end{equation}
Here, the superscript $i$ stands for the spatial components of the corresponding four-vector, and $\nabla$ is the flat  spatial  vector differential operator. Besides, 
$B^i$ represents the spatial components of the magnetic field, defined in general as $B^\mu=n_\rho \epsilon^{\rho\mu\nu\alpha} F_{\nu\alpha}/2$, where $\epsilon^{\rho\mu\nu\alpha}$ is the totally antisymmetric tensor, with
 the four-vector  $n_\mu = (1, 0, 0, 0)$ \cite{faz}. Similarly, for the positron fluid, we find
\begin{equation}
\nabla\times\left(f U_p^i \right)=\frac{e a}{m} B^i\, .
\end{equation}
Using these solutions in Maxwell equations \eqref{ecMAxwellll}, we can find an electromagnetic plasma wave equation for the dynamics of the vector magnetic field {\bf B} in a general cosmological scenario \cite{faz}
\begin{equation}
    \frac{1}{a^2}\frac{\partial}{\partial t}\left(a \frac{\partial}{\partial t}\left(a^3 {\bf B}\right)\right)-\nabla^2 {\bf B}+\omega_p^2\, V(a){\bf B}=0\, .
    \label{waveeq1}
\end{equation}
The constant plasma frequency of this plasma system is $\omega_p=\sqrt{8 \pi e^2 n_0/(m f_0)}$, where   $n_0$ is the rest-frame electron (and positron) density at the current  time of the Universe $t_0$, and $f_0$ is the current value of the function $f$. Lastly, the function $V$ depends on the scale factor $a$, and it is given by \cite{faz}
\begin{equation}
    V(a)= \frac{f_0}{f(a)}\frac{n(a)}{n_0}a^2=\frac{f_0}{a\, f(a)}\, ,
    \label{potentialV}
\end{equation}
where we have considered that the number density of plasma particles decreases
with the volume of the Universe as $n(a) = n_0/a^3$ \cite{holcomb}. The function $V$ acts as a potential, and it
measures the effective cosmological   response of the plasma to the passing of an electromagnetic wave. It depends also on whether the plasma is hot or cold (through $f$), as we see below. In a nonexpanding Universe, $a=1$, $f=f_0$,  $V=1$, and the solutions of Eq.~\eqref{waveeq1} are simple plane waves, with their corresponding dispersion relation.

In terms of the conformal time
\begin{equation}
    \tau=\int\frac{dt}{a}\, ,
\label{cosmotime}
\end{equation}
the magnetic field can be studied in the form
\begin{equation}
{\bf B}(\tau,{\bf x})=\frac{1}{a^3} {\bf b}(\tau)\exp(i {\bf k}\cdot{\bf x})\, ,
\label{formmagnetic}
\end{equation}
where ${\bf k}$ is a wavenumber vector.
The form of this ansatz solution, in terms of separation of variables, is determined by the functionality of the potential, depending only on the conformal time $\tau$. Also, we have taken into account the cosmological volumetric decay of the field as the Universe expands.
Thus, wave equation \eqref{waveeq1} reduces now to
 \begin{equation}
     \frac{\partial^2{\bf b}}{\partial \tau^2}+\left[k^2+\omega_p^2\, V(\tau)\right]{\bf b}=0\, ,
    \label{waveeq2}
\end{equation}
where  the potential $V$ must be put in terms of $\tau$ for the appropriated cosmological scenario, and $k^2={\bf k}\cdot{\bf k}$. The  field ${\bf b}$ is the effective cosmological magnetic field that depends on the kind of evolution of the Universe.
All solutions \eqref{formmagnetic} must be in agreement with the constraint $\nabla\cdot{\bf B}=0$.

For a cosmological model with a cosmological constant as a dark energy component of the Universe, with scale factor \eqref{scalefactor}, we find, from Eq.~\eqref{cosmotime}, that the conformal time is given by
\begin{equation}
    \tau=\frac{e^{H_0 t_0}}{H_0}\left(1-e^{-H_0 t}\right)\, .
\end{equation}
In this model, we are considering this cosmological scenario to evolve from $t=0$ to $t\rightarrow\infty$. Then, we have $0\leq \tau \leq e^{H_0 t_0}/H_0$.
In this way, we can put the scale factor in terms of the conformal time as
\begin{equation}
    a(\tau)=\left(e^{H_0 t_0}-H_0\tau\right)^{-1}\, .
\end{equation}

Electromagnetic plasma waves can be solved perturbatively in general \cite{holcomb}. However, for the cosmological model \eqref{scalefactor},
 exact solutions of Eq.~\eqref{waveeq2} can be found. In the following, we show two of these solutions, for   cold and hot plasma cases. These two plasma temperature limits change the form of the potential $V$.

\subsection{Cold plasma scenario}

In this case, when $T\ll m$, we have  \cite{fazmunoz}
\begin{equation}
    f(a)\approx 1+\frac{5 T(a)}{2 m}\, .
\end{equation}
Besides, the cosmological  plasma temperature  decays as \cite{Dodin}
\begin{equation}
    T(a)\approx \frac{T_0}{a^2}\, ,
\end{equation}
where $T_0$
is the actual plasma temperature. Then, $f_0/f\approx 1-5 T_0/(2m a^2)$, and the
potential \eqref{potentialV}
becomes simply
\begin{equation}
    V(a)\approx \frac{1}{a}=V(\tau)=e^{H_0 t_0}-H_0\tau\, .
\end{equation}
Using this, Eq.~\eqref{waveeq2} becomes
 \begin{equation}
     \frac{\partial^2{\bf b}}{\partial \tau^2}+\left[k^2-\omega_p^2 H_0 \tau+\omega_p^2e^{H_0 t_0}\right]{\bf b}=0\, ,
\end{equation}
which is  the equation for Airy functions. Thereby, 
we readily find the solution of the effective electromagnetic field  as
\begin{equation}
    b(\tau)={\mbox{Ai}}\left(\left(\omega_p^2H_0\right)^{1/3}\tau-\left(\omega_p^2 H_0\right)^{-2/3}\left(\omega_p^2e^{H_0 t_0}+k^2\right) \right)\, ,
    \label{airysocold}
\end{equation}
being $b$ the magnitude of ${\bf b}$, and ${\mbox{Ai}}$ stands for the Airy function. This problem is, therefore, the cosmological time-domain equivalent to a quantum particle in a non-relativistic gravitational potential \cite{Kajari}.

This electromagnetic wave has an amplitude that oscillates in a standing form with the Airy functionality given by
\eqref{airysocold}. Thus, because of this, as $\tau\rightarrow \exp(H_0 t_0)/H_0$, the amplitude of the wave grows to reach a maximum amplitud and then it  gets damped.
We can determine the amount of time in which this solution oscillates. Using the properties of Airy functions, we find that these waves oscillate  a time $\Delta\tau$, given by
\begin{equation}
    \Delta\tau=\left(H_0\omega_p^2\right)^{-1/3}{\mbox{Ai}}_{M1}+\left(H_0\omega_p^2\right)^{-1}\left(\omega_p^2  \exp(H_0 t_0)+k^2  \right)\, ,
\label{maximumtime}
\end{equation}
where
${\mbox{Ai}}_{M1}\approx  -1.01879$  determines first and largest maximum of the Airy function.
Before this time,  for any $\tau$ such that $0\leq\tau \leq \Delta \tau$, the oscillations always grow.
After this, for times $\tau$ such that $\Delta\tau<\tau\leq e^{H_0t_0}/H_0$, the wave decays  due to its Airy functionality. 
The decaying of this mode  certainly occurs if 
\begin{equation}
k^2\ll (-{\mbox{Ai}}_{M1})(\omega_p^2 H_0)^{2/3} \, .
\end{equation}
  Thereby, in this case, the wavepacket has a non-zero intial value at $\tau=0$, growing in amplitude till reaches a maximum, to later
decrease its amplitude as  time  approaches to $\tau=\exp(H_0 t_0)/H_0$.

A different case can be studied when the magnetic field is required to satisfy initial conditions. Because of the dark energy cosmological constant Universe model starts at $t=0$, then we can choose that the electromagnetic wave  vanishes in $\tau=0$ (equivalent to $t=0$). Due to this, we find that
\begin{equation}
    \left(H_0\omega_p^2\right)^{-2/3}\left(\omega_p^2\, e^{H_0 t_0}+k^2\right)=-{\mbox{Ai}}_{n}\, ,
\end{equation}
where ${\mbox{Ai}}_{n}<0$ is any of the $n$th zero of the Airy function ($n= 1, 2, 3, . . .$). This implies that the existence of this electromagnetic wave is only possible if its wavelength $\lambda=2\pi/k$ is  discretized with the form
\begin{equation}
    \lambda_n=2\pi\left(-\left(H_0\omega_p^2\right)^{2/3}{\mbox{Ai}}_{n}-\omega_p^2\, e^{H_0 t_0}\right)^{-1/2}\, ,
\end{equation}
depending, therefore, on the response frequency  of the plasma medium $\omega_p$ to its propagation, on $H_0$, and  on the chosen initial
${\mbox{Ai}}_n$. Therefore, in this case, the electromagnetic wavepacket acquire a fixed value (depending on $n$) at the final time $\exp(H_0 t_0)/H_0$ of its evolution.

\subsection{Hot plasma scenario}

In contrast  to the above, in the case of $T\gg m$, we have \cite{fazmunoz}
\begin{equation}
    f(a)\approx \frac{4 T(a)}{m}+\frac{m}{2T(a)}\, .
\end{equation}
However, in this case \cite{Dodin}
\begin{equation}
    T(a)\approx \frac{T_0}{a}\, . 
\end{equation}
Thus, $f/f_0\approx 1/a+m^2 a/(8 T_0^2)$, and 
\begin{equation}
    V(a)\approx 1-\frac{m^2a^2}{8T_0^2}=V(\tau)=1-\frac{m^2}{8T_0^2}\left(e^{H_0 t_0}-H_0\tau\right)^{-2}\, .
\end{equation}
This potential can be used in Eq.~\eqref{waveeq2}, which becomes
 \begin{equation}
     \frac{\partial^2{\bf b}}{\partial \tau^2}+\left[k^2+\omega_p^2 -\frac{m^2 \omega_p^2}{8T_0^2}\left(e^{H_0 t_0}-H_0\tau\right)^{-2}\right]{\bf b}=0\, .
\label{eq27hot}
\end{equation}
By noting that the effective time in this equation is $e^{H_0 t_0}/H_0-\tau$,  we get that this equation is the one satisfied for Bessel functions. Therefore, the electromagnetic wave solution propagates in the form
\begin{equation}
    b(\tau)=\sqrt{e^{H_0 t_0}-H_0\tau}\, {{J}}_\beta\left(\sqrt{k^2+\omega_p^2}\left(\frac{1}{H_0}e^{H_0 t_0}-\tau \right) \right)\, ,
\label{solbessel}
\end{equation}
where ${{J}}_\beta$ is the Bessel function of the first kind of order $\beta$, with
\begin{equation}
    \beta=\frac{1}{2}\sqrt{1+\frac{m^2\omega_p^2}{2T_0^2 H_0^2}}\, .
\end{equation}
This solution is the  physically acceptable one for a dark energy Universe (due to a cosmological constant) if we require that the electromagnetic wave vanishes at $t\rightarrow\infty$, or $\tau=e^{H_0 t_0}/H_0$.
Therefore, this solution implies, similar to the cold plasma case, that due to  the initial vanishing condition for the wave in $\tau=0$ (or $t=0$),  we have
\begin{equation}
    \frac{\sqrt{k^2+\omega_p^2}}{H_0}e^{H_0 t_0}=j_{\beta,n}\, ,
\end{equation}
where $j_{\beta,n}$ ($n= 1, 2, 3, . . .$) are any of the $n$th zero of the Bessel function of order $\beta$. Thus, for the hot case, the wavelength of this wave becomes also discrete, with the spectrum
\begin{equation}
    \lambda_n=2\pi \left(H_0^2 j_{\beta,n}^2 e^{-2 H_0t_0}-\omega_p^2\right)^{-1/2}\, .
\end{equation}
Only waves with these wavelengths are allowed to exist exactly in a hot plasma in this dark energy cosmology, under the discussed conditions. This solution is possible  under  a condition on the plasma frequency $\omega_p$, which must have an upper limit.

On the contrary to the above solution, if we relax the vanishing condition of the wave at initial and final times, then another solution of 
Eq.~\eqref{eq27hot} is possible. This is
\begin{equation}
    b(\tau)=\sqrt{e^{H_0 t_0}-H_0\tau}\, {\mbox{H}}^{(1)}_\beta\left(\sqrt{k^2+\omega_p^2}\left(\frac{1}{H_0}e^{H_0 t_0}-\tau \right) \right)\, ,
\label{soluhankel}
\end{equation}
where $ {\mbox{H}}^{(1)}_\beta$ is the Hankel function of the first kind of order $\beta$.
This solution diverges when $t\rightarrow \infty$, as the argument of the Hankel function goes to zero. Also, in general, the order of the Hankel function is not $1/2$, because of the hot plasma temperature. Thus, this result is a generalization of those found, for example, in Ref.~\cite{Banerjee}.  

\section{Discussion}

Usually, plasma wave dynamics in cosmology are solved in a perturbative fashion. However, in this work, we show that in the case of a dark energy cosmology, triggered by a cosmological constant, the electromagnetic plasma waves can be solved exactly for a hot and a cold plasma.

These electromagnetic plasma waves evolve according to a time-dependent potential \eqref{potentialV} that
effectively modifies the frequency response of the plasma through the passing of the electromagnetic wave. Remarkably, cold and hot plasmas produce very different outcomes on the type of electromagnetic waves that can exist in the dark energy cosmology. The vanishing initial and final conditions (in time) force those solutions to develop in a way similar to standing waves.
In the hot plasma scenario, those conditions play the role of  absorbers, and then the presented solutions can be put in the context of time-symmetrical electromagnetic solutions of the Wheeler-Feynman absorber theory \cite{hogarth}. Those conditions also produce a classical discretization on the wavelength of the wave, which
depends on the interplay of the dark energy features and the response of the plasma. Notice that this behavior is not possible for simple light propagating in cosmology [when $\omega_p=0$ in Eq.~\eqref{waveeq2}].

For the cold plasma scenario, the Airy form of the solution forces to have a discrete spectrum of plasma wave wavelengths when the wave is required to vanish at initial times.
The wavelengths depend on $H_0$ and on the plasma frequency.
On the other hand, the interaction of this cosmology and the cold plasma impose a growth on the oscillations of the
electromagnetic wave, reaching a maximum on time  \eqref{maximumtime}. After that, this exact mode gets mainly damped for larger times. 
This mode can be quickly damped if $\Delta\tau\ll e^{H_0t_0}/H_0$, which imposes a condition on the
cosmological constant $\Lambda$ (or $H_0$) as $ k^3/\omega_p^2\ll H_0$. This condition implies that, for given  $H_0$, large wavelength plasma modes probably cannot be observed  in a dark energy cosmology.

On the contrary, a hot plasma allows the electromagnetic wave to oscillate in the form \eqref{solbessel} to $t\rightarrow\infty$, where it vanishes. Anew, the dark energy cosmology forces the discretization of the wavelengths of any possible modes that can exactly exist. They are again imposed by the initial conditions of the Universe. Also, this discretization
depends on the interplay of the plasma (through the plasma frequency) and the dark energy (through $H_0$). This oscillation does not get damped, as in the cold case, but only it decreases for larger times, vanishing at initial and final times.
If those conditions are not taken into consideration, solution \eqref{soluhankel} is obtained, which represents a distinct form of plane wave   oscillation. This is best represented in the very hot case, when $m\omega_p\ll T_0 H_0$ and $\beta\approx 1/2$. In this limit,
solution \eqref{soluhankel} becomes
\begin{equation}
b(\tau)\approx {\frac{\sqrt{2 H_0/\pi}}{(k^2+\omega_p^2)^{1/4} }}\exp\left( i\sqrt{k^2+\omega_p^2}\left(\frac{1}{H_0}e^{H_0 t_0}-\tau \right) \right)\, ,
\end{equation}
being a plane electromagnetic plasma wave with a time-dependent frequency (in cosmological time $t$)
given by $\omega(t)={\sqrt{k^2+\omega_p^2}}/{a}$,
which takes into account the cosmological redshift.
 
 Finally, both (cold and hot) solutions presented in this work are possible only when the plasma system does not affect the evolution of the dark energy Universe. If these solutions remain to be qualitatively similar in the case that energy density of the plasma contributes to the Einstein equations, is an open question to be studied.

\begin{acknowledgements}
FAA thanks to FONDECYT grant No. 1230094 that partially supported this work. 
 \end{acknowledgements}

\end{document}